\documentclass[pra,superscriptaddress,showpacs,preprintnumbers,
 eqsecnum,amsmath,amssymb]{revtex4}

\usepackage{color}

\begin{document}

\title{Classical and Quantal State Reconstruction }

\author{M. Revzen}
\affiliation {Department of Physics, Technion - Israel Institute of Technology, Haifa
32000, Israel}
\date{\today}

\begin{abstract}
 Analysis of state reconstruction  both classical and quantum mechanical on equal footing
is outlined. The meaning of "mutual unbiased bases" (MUB) of Hilbert spaces is explained in
detail. An alternative quantum state reconstruction, that utilizes mutual unbiased bases
(MUB), is given. The MUB approach is then used for state reconstruction in a finite, d,
dimensional Hilbert spaces.
\end{abstract}


\maketitle


\section {Introduction}

The position operator, $\hat{x}$ and the momentum operator $\hat{p}$ form a complete
operator basis \cite{schwinger}, i.e. an operator that commutes with both must be a scalar.
In view of this one might expect that the reply to W. Pauli query \cite{schleich}, viz: can
one construct a wave function, $\psi(x)$, amplitude and phase, from the probabilities for
(all) x and p, would be in the affirmative. i.e. that $|\psi(x)|^2$ and $|\psi(p)|^2$ have
sufficient information to deduce $\psi(x)$. However this is {\it not} sufficient. Indeed
what is required to reconstruct a state is the issue dealt with in this pedagogical
article.  In other words we will see that measuring (assuming unlimited ensemble of
$|\psi\rangle$) the position $\langle x|\psi\rangle\langle \psi|x\rangle$ and the momentum
$\langle p|\psi\rangle\langle \psi|p\rangle$ would {\it not} allow the reconstruction of
the state. To reconstruct the state one requires the probability distributions of all the
so called mutually unbiased operators (these will be defined below) - this set, of which
the operators $\hat{x}$ and $\hat{p}$, are members will now be considered.\\
The conceptual idea of mutual unbiased operators was introduced by Schwinger
\cite{schwinger} when he considered vectorial bases for Hilbert spaces which exhibit
"maximal degree of incompatibility" - the eigenvectors of $\hat{x}$ - $|x\rangle$ and of
$\hat{p}$, $|p\rangle$ are example of such bases. The information theoretical oriented
appellation  "mutual unbiased bases" (MUB) were introduced by
\cite{wootters1,wootters2,wootters3} with the following formal definition:  Two orthonormal
vector bases, $ {\cal B}_{1},\;{\cal B}_{2}$, are said to be mutually unbiased (MUB) if and
only if  ($ {\cal B}_{1} \ne {\cal B}_{2}$)
\begin{equation}
\forall\; |u_1\rangle,\;|u_2\rangle\;\epsilon\; {\cal B}_{1},\;{\cal B}_{2}\;\;resp.
\;\; |\langle u_1|u_2 \rangle |=constant,
\end{equation}
 i.e. the absolute value of the scalar product of vectors from different bases is
 independent of the vectorial labels within either basis. This implies that if the
 state vector is measured to be in one of the states e.g. $|u_1\rangle$ of the base
${\cal B}_{1}$, it is equally likely to be in any of the states $|u_2\rangle$ of the base
${\cal B}_2$. (The value of $|\langle u_1|u_2 \rangle |$ may depend on the bases, ${\cal
B}_{1},\;{\cal B}_{2}$ as indeed is the case for the continuous  dimensionality - $d
\rightarrow \infty$, - limit.) MUB were found of interest in several fields. Thus the ideas
are useful in variety of cryptographic protocols (e.g. \cite{ekert}), signal analysis
\cite{schroeder}. They are of particular interest in quantum state tomography, \cite{ulf}
where quadrature \cite{walls} observations allow the construction of Wigner function of the
state under study. The relation of MUB to Wigner function is extensively studied by
Wootters and co-workers \cite{wootters1,wootters2,wootters3}. Much of these works were
devoted to finite dimensional Hilbert spaces. Here it was shown that a d dimensional
Hilbert space can accommodate at most d+1 MUB and the value of $|\langle u_1|u_2 \rangle |$
is $\frac {1}{\sqrt d}$ \cite {ivanovich,wootters1,tal,bengtsson,amir}. The finite
dimensional theory is intriguingly connected to algebraic field theory
\cite{ivanovich,wootters1, wootters2,tal,vourdas1,vourdas2,klimov1}; connections to other
sophisticated mathematical notions are given in
\cite{wootters4,planat1,planat2,combescure,klimov2} and summarized in \cite{bengtsson}.
Extensive infinite dimensional studies are given in
\cite{wootters1,wootters2,wootters3,stefan}. We will argue that the information required to reconstruct a
quantum state may be gained by measuring all the mutually unbiased operators.\\
The problem of state reconstruction has, of course, its classical analogue. Within
classical physics we may refer to the density in phase space, $\rho(x,p)$, of a physical
system as the state (of the system). In the following we will study how, i.e. what kind of
measurements, will suffice to reconstruct such a state. The methodology will be seen to
based on the inverse Radon transform \cite{ulf}. Then, utilizing the quasi probability
nature of the Wigner representative function  $W_{\rho}(x,p)$  \cite{ulf,schleich} of the
quantum state operator $\hat{\rho}$, \cite{feynman}, we show that reconstructing the Wigner
function is closely analogous to the reconstruction of classical phase space distribution
function. This is the content of Section II.\\
 Section III contains an alternative derivation of the
quantum state reconstruction which is based on MUB. This approach does not require an
explicit use of Radon transform. In Section IV we give an MUB based state reconstruction
for finite, d, dimensional cases with d a prime number. This approach does not involve
reconstruction of the Wigner representative function whose finite dimensional version is
rather challenging \cite{ulf2,wootters1}. The last section contains summary and discussion.

\section{Classical-Quantum State Reconstructions Analogy}

The classical state, $\rho(x,p)$, is the phase space distribution function. In terms  of
this state the classical probability (density - to be understood henceforth),
$\rho(x',p')$, for the particle position to be at the local $x=x'$ and the momentum of
$p=p'$ is

\begin{equation}
\rho(x',p')=\int dxdp \rho(x,p)\delta(x-x')\delta(p-p'),
\end{equation}
 The Fourier transform of the distribution, $F(a,b;\rho)$, is

\begin{equation}\label{ab}
F(a,b;\rho)=\int \frac{dxdp}{2\pi}e^{iax}e^{ibp}\rho(x,p).
\end{equation}

Thus,

\begin{equation}
\rho(x',p')=\int\frac{dadb}{2\pi}e^{-iax'}e^{-ibp'}F(a,b;\rho) =\int
dxdp\rho(x,p)\delta(x-x')\delta(p-p').
\end{equation}

The marginal probability for the particle being at $x=x'$ while being completely ignorant
of its momenta, is
\begin{equation}
\tilde{\rho}(x',0)=\int dxdp \rho(x,p)\delta(x-x').
\end{equation}
(The entry label 0 will become meaningful below.) Correspondingly the marginal distribution
that the particle's phase space coordinate be $x'=X_{\theta}\equiv xC+pS;$
 $(C=cos\theta,\;S=sin\theta)$ is given by the following transform,
\begin{equation}\label{marg1}
\tilde{\rho}(x',\theta)=\int dxdp \rho(x,p)\delta(x'-xC-pS).
\end{equation}
This transformation is known as Radon transformation, \cite{ulf}. $\tilde{\rho}(x',\theta)$
is a directly measurable quantity \cite{ulf}.Before considering the inversion of this, i.e.
getting $\rho(x,p)$ in terms of $\tilde{\rho}(x',\theta)$, we note that such, (Eq.(
\ref{marg1})), marginal distribution implies complete lack of knowledge of the odds for the
realization of phase space coordinate $p'=P_{\theta}\equiv -xS+pC$. The proof is as
follows: define the $\theta$ depending variables
\begin{equation}
x_{\theta}=Cx+Sp,\;\;p_{\theta}=-Sx+Cp.
\end{equation}
Now one readily verifies that
\begin{equation}
\tilde{\rho}(x',\theta)=\int dxdp \rho(x,p)\delta(x'-xC-pS)=\int dx_{\theta}dp_{\theta}
\rho(x(x_{\theta},p_{\theta}),p(x_{\theta},p_{\theta}))\delta(x'-x_{\theta}).\;QED
\end{equation}

The inversion of the formula proceeds as follows. Expressing the delta function in Eq.(
\ref{marg1}) in terms of its Fourier transform

\begin{eqnarray}\label{rC}
\tilde{\rho}(x',\theta) &=&\int dr e^{-ix'r}\tilde{F}(r;\theta,\rho),\nonumber\\
\tilde{F}(r;\theta,\rho)&=& \int dxdp\rho(x,p)e^{iCrx}e^{iSrp},\;i.e. \nonumber\\
\tilde{F}(r;\theta,\rho)&=&F(a=rC,b=rS;\rho),\;and, \nonumber\\
\tilde{F}(r;\theta,\rho)&=&\int \frac{dx'}{2\pi} e^{ix'r}\tilde{\rho}(x',\theta).
\end{eqnarray}
This relates the measurable $\tilde{\rho}(x',\theta)$ to $\tilde{F}(r;\theta,\rho)$. The
inverse Radon transform is obtained as follows
\begin{equation}\label{radon}
\rho(x,p)=\int \frac{dadb}{2\pi}
e^{-iax}e^{-ibp}F(a,b;\rho)=\int_0^{\infty}rdr\int_0^{2\pi}\frac{d\theta}{2\pi}
e^{-irCx}e^{-irSp}\tilde{F}(r;\theta,\rho).
\end{equation}
Performing the r integration and noting that
\begin{equation}\label{-x,theta}
\tilde{\rho}(-x',\theta)=\tilde{\rho}(x',\theta+\pi)
\end{equation}
 we get
\begin{equation}
\rho(x,p)=-\frac{\mathcal{P}}{2\pi^2}\int_0^{\pi}\int_{-\infty}^{\infty}\frac{\tilde{\rho}(x',\theta)dx'd\theta}
{(xC+pS-x')^2}.
\end{equation}
This formula reconstructs, via the inverse Radon transform, the full (classical)
distribution, $\rho(x,p)$ from marginals, "sliced" distributions,
$\tilde{\rho}(x',\theta),$ which may be
obtained by direct measurements. As noted above, Eq. (\ref{marg1}) is an example of (the direct) Radon transform \cite{ulf}.\\

The quantum reconstruction utilizes the close analogy between the classical phase space
distribution function and the Wigner representative function $W_{\rho}(x,p),$
\cite{ulf,schleich,wigner,mello1,mello2} of the density operator $\hat{\rho}$,
\begin{equation}
W_{\rho}(x,p)=\int\frac{dy}{2\pi}e^{ipy}\langle
x-\frac{y}{2}|\hat{\rho}|x+\frac{y}{2}\rangle.
\end{equation}
Consider the operator $\hat{X}_{\theta}\equiv \hat{x}C+\hat{p}S$ and its eigenstates,
\begin{equation}\label{basis}
\hat{X}_{\theta}|x',\theta\rangle=x'|x',\theta\rangle,
\end{equation}
these form a complete orthonormal basis \cite{wootters1}. Now the probability of finding
the particle in the state $|x',\theta\rangle$ given that its (the particle's) state is
$\hat{\rho}$ is,
\begin{eqnarray}\label{meas}
\tilde{\rho}_{Q}(x',\theta)&=&Tr{|x',\theta \rangle \langle x',\theta|\hat{\rho}}\nonumber \\
                       &=&\langle x',\theta|\hat{\rho}|x',\theta\rangle.
\end{eqnarray}
The quantity $\langle x',\theta|\hat{\rho}|x',\theta\rangle$ is directly measurable
 \cite{ulf}. Transcribing this to the Wigner representative function,
\begin{equation}\label{trace}
\tilde{\rho}_{Q}(x',\theta)=\int\frac{dxdp}{2\pi}W_{|x',\theta\rangle}(x,p)W_{\rho}(x,p).
\end{equation}
Noting that (the proof is given in the next section)

\begin{equation}\label{ls}
\langle x'|x;\theta \rangle=\frac{1}{\sqrt{2\pi
|S|}}e^{-\frac{i}{2sin\theta}\left([x^2+{x'}^2]C-2xx'\right)}.
\end{equation}
We evaluate,
\begin{equation}
W_{|x',\theta \rangle}(x,p)=\int\frac{d}{2\pi}y e^{ipy}\langle x-\frac{y}{2}||x',\theta
\rangle\langle x',\theta|x+\frac{y}{2}\rangle=\delta(x'-xC-pS).
\end{equation}
Eq.(\ref{meas}) now reads,
\begin{equation}\label{trace}
\tilde{\rho}_{Q}(x',\theta)=\int\frac{dxdp}{2\pi}\delta(x'-xC-pS)W_{\rho}(x,p).
\end{equation}
Which, using the analysis given above for the classical case, Eq.(\ref{rC},\ref{radon}),
gives, \cite{ulf}
\begin{equation}\label{ulf}
W_{\rho}(x,p)=-\frac{\cal{P}}{2\pi^2}\int_0^{\pi}\int_{-\infty}^{+\infty}
\frac{\tilde{\rho}_{Q}(x',\theta)dx'd\theta}{(xC+pS-x')^2}.
\end{equation}
 Thus measuring
$\langle x',\theta|\hat{\rho}|x',\theta\rangle$ i.e. $\tilde{\rho}_{Q}(x,\theta)$ (cf.
Eq.(\ref{meas}), yields the Wigner representative function for the state $\hat{\rho}$
thereby reconstructing the state since $W_{\rho}(x,p)$ is a faithful
representative of the quantum state \cite{ulf,schleich,wigner}.\\

\section{State Reconstruction via Mutual Unbiased Bases (MUB)}

We now outline an alternative state reconstruction strategy which is based on MUB. We begin
with a brief review of a recently \cite{wootters3,revzen} MUB construction: The complete
orthonormal basis \{$|x',\theta\rangle$\} as defined above, Eq.(\ref{basis}), constitute a
set of bases labelled by $\theta$, each complete and orthonormal.. We shall now argue that
these are MUB. We have, as can be checked directly \cite{ulf,revzen},
\begin{equation}
\hat{X}_{\theta}=U^{\dagger}(\theta)\hat{x}U(\theta),
\end{equation}
with,
\begin{equation}\label{zpe}
\hat{a}\equiv \frac{1}{\sqrt 2}(\hat {x}+i\hat{p});\;\hat{a}^{\dagger} \equiv
\frac{1}{\sqrt 2} (\hat {x}-i\hat{p});\;U(\theta) \equiv e^{-i\theta
\hat{a}^{\dagger}\hat{a}}.
\end{equation}

Returning to Eq.(\ref{basis}), we have that solutions of the equation may be written in
terms of $\{|x\rangle\}$, the eigenvectors of the position operator, as
\begin{equation}\label{prels}
|x;\theta\rangle=U^{\dagger}(\theta)|x\rangle.
\end{equation}
This defines our phase choice. Thus we may read off the x-representative solutions
\cite{larry,revzen},
\begin{equation}\label{ls}
\langle x'|x;\theta \rangle=\frac{1}{\sqrt{2\pi
|S|}}e^{-\frac{i}{2S}\left([x^2+{x'}^2]C-2xx'\right)}.
\end{equation}
This can be verified directly as the solution of
\begin{equation}
\big(Cx-iS\frac{\partial}{\partial x}\big)\langle x'|x,\theta\rangle=x\langle
x'|x,\theta\rangle.
\end{equation}

 With the phase choice above the state,
$x\leftrightarrow x'$, is symmetric for $x\leftrightarrow x'$, a property that facilitate
several calculations. We can now verify that the bases $\{|x;\theta\rangle\}$ and
$\{|x';\theta'\rangle\}$ with $\theta \neq \theta'$ are MUB: thus \cite{amir,revzen}
\begin{equation}
\langle x';\theta'|x;\theta\rangle=\langle x'|U^{\dagger}(\theta-\theta') x\rangle|=\langle
x|x;\theta-\theta'\rangle=\frac{1}{\sqrt{2\pi |S(\theta,\theta')|}}
e^{-\frac{i}{2S(\theta,\theta')}\left([x^2+{x'}^2]C(\theta,\theta')-2xx'\right)}.
\end{equation}
Here $S(\theta,\theta')\equiv\sin(\theta-\theta')$ and
$C(\theta,\theta')\equiv\cos(\theta-\theta')$. Indeed,
\begin{equation}
|\langle x';\theta'|x;\theta\rangle|=\frac{1}{\sqrt{2\pi |S(\theta,\theta')|}}\;,
\end{equation}
the number $|\langle x';\theta'|x;\theta\rangle|$ is independent of the vectorial labels $x$ and $x'$,
 i.e. the bases $\{|x;\theta\rangle\}$ with distinct $\theta$ are MUB.
We further note that
\begin{eqnarray}
 \lim_{\theta \rightarrow 0} \langle x|x';\theta \rangle &\rightarrow& \delta(x-x') \nonumber \\
 \lim_{\theta \rightarrow \frac{\pi}{2}}\langle x|x';\theta \rangle &\rightarrow&
\frac{e^{ixx'}}{\sqrt{2 \pi}}.
\end{eqnarray}
This concurs with the observation that, (c.f., Eq. (2.4)), in the first limit $|x;0\rangle$
is the eigenfunction of $\hat{x}$ whose x representative is $\delta(x-x')$. In the second
limit, $|x';\frac{\pi}{2}\rangle$ is the eigenfunction of the momentum operator ($\hat{p}$)
thus the limit is simply a Fourier representative expressing the momentum in the x
representation.\\

We now derive an explicit tomographic formula for the density operator i.e. the state is
reconstructed via expressing the state, $\langle x|\rho|x'\rangle$ in terms of measurable
$\langle x,\theta|\rho|x,\theta \rangle$. To this end we note that the set
\begin{equation}
e^{ia\hat{x}}e^{ib\hat{p}},\;\;-\infty\le a,b\le +\infty,
\end{equation}
is a complete and orthogonal operator basis \cite{weyl}
\begin{equation}
\int
\frac{dadb}{2\pi}tr\big(e^{ia\hat{x}}e^{ib\hat{p}}[e^{ia'\hat{x}}e^{ib'\hat{p}}]^{\dagger}\big)=\delta(a-a')\delta(b-b').
\end{equation}
Thus we may expand,

\begin{equation}\label{rhoab}
\hat{\rho}\;=\;\frac{1}{2\pi}\int dadb (tr\rho e^{ia \hat{x}}e^{ib \hat{p}})(e^{ia
\hat{x}}e^{ib \hat{p}})^{\dagger}.
\end{equation}
Defining $a=rcos\theta,\;b=rsin\theta$ and using the equality,
$$e^{ia \hat{x}}e^{ib \hat{p}}=e^{(i\frac {r^2sin
{2\theta}}{4})}e^{i(cos \theta \hat{x}+sin \theta \hat{p})r},$$ we may rewrite Eq.
(\ref{rhoab}) as,
\begin{equation}
\hat{\rho}=\int_{0}^{\infty} \int_{-\pi}^{\pi}\frac{rdr d\theta}{2\pi} (tr\rho e^{(i\frac
{r^2sin {2\theta}}{4})}e^{i(cos \theta \hat{x}+sin \theta \hat{p})r})\left(e^{(i\frac
{r^2sin2\theta}{4})}e^{i(cos \theta \hat{x}+sin \theta \hat{p})r}\right)^{\dagger}.
\end{equation}
Utilizing the spectral representation,
\begin{equation}
e^{i(cos \theta \hat{x}+sin \theta \hat{p})r}=\int dc|c,\theta\rangle e^{irc}\langle
c,\theta|,
\end{equation}

and evaluating the trace via the orthonormal bases $\{|c;\theta\rangle\}$, this expression
becomes
\begin{equation}
\int dcdc'\int_{0}^{\infty} \int_{-\pi}^{\pi}\frac{rdr d\theta}{2\pi} \langle
c;\theta|\rho|c;\theta\rangle e^{irc}|c';\theta\rangle\langle c';\theta|e^{-ic'r}.
\end{equation}
Evaluating  the integral
\begin{equation}
\int_{0}^{\infty} rdr e^{ir(c-c')}=-i\frac{d}{dc}\int_{0}^{\infty}dr e^{ir(c-c')}=-{\cal
P}\frac{1}{(c-c')^2},
\end{equation}
and inserting it in the above expression we get our tomographic formula (the bounds on the
$\theta$ integration is justified via Eq.(\ref{-x,theta})):
\begin{equation}\label{tom}
\hat{\rho}=-\frac{\cal P}{\pi}
\int_{-\infty}^{\infty}dcdc'\int_{0}^{\pi}\frac{d\theta}{(c-c')^2} \langle
c;\theta|\rho|c;\theta\rangle|c';\theta\rangle\langle c';\theta|.
\end{equation}

\noindent  Thus the density matrix, $\langle x|\rho|x' \rangle$, is given in terms
of the observables  $\langle c;\theta|\rho|c;\theta\rangle$:
\begin{equation}
\langle x|\rho|x' \rangle=-\frac{\cal P}{\pi}
\int_{-\infty}^{\infty}dcdc'\int_{0}^{\pi}\frac{d\theta}{(c-c')^2} \langle
c;\theta|\rho|c;\theta\rangle
\frac{e^{\frac{i}{2S}\left((x^2-x'^2)C-2(x-x')c'\right)}}{2\pi S}.
\end{equation}
This completes our state reconstruction via MUB: one may construct $\langle x|\rho|x'
\rangle$ in terms of the measurables  $\langle c;\theta|\rho|c;\theta\rangle.$  The
expression is of course equivalent to the one given in terms of Wigner function
 \cite{ulf}: Evaluating the Wigner function in Eq. (\ref{tom}) gives,
\begin{equation}
W_{\rho}(q,p)=-\frac{\cal P}{2\pi^2}\int d\theta dc\frac{\langle
c;\theta|\rho|c;\theta\rangle}{(qC+pS -c)^2}.
\end{equation}
Where the Wigner function is given in terms of measurements that give the
probabilities $\langle c;\theta|\rho|c;\theta\rangle$, conforming with \cite{ulf}.

\section{State Reconstruction in Finite Dimensions}

The difference in the two strategies discussed above for the continuous phase space
(Hilbert space dimensionality $d \rightarrow \infty)$ involve, in essence, the benefit of
two ways of handling the complete orthonormal operators $e^{ia \hat{x}}e^{ib \hat{p}}$.
Thus when viewing them as Fourier transformation \cite{ulf} we are led to the Wigner
function via the Radon transformation. If, on the other hand, we consider the spectral
representation (note $a=rC, b=rS,\;c=cos\theta,\;S=sin\theta$),
\begin{equation}
e^{ia \hat{x}+ib \hat{p}}=e^{ir(\hat{x}C+\hat{p}S)}=\int dx'|x',\theta \rangle
e^{irx'}\langle x',\theta|,
\end{equation}
we are headed for state reconstruction via MUB. This approach will now be broadened
to accommodate the finite ,d, dimensional Hilbert space state reconstruction \cite{amir}.\\

Schwinger \cite{schwinger} noted that the physics of finite dimensional, d, Hilbert space
is accountable via two unitary operators, $X$, and $Z$. Thus if we label the d distinct
states, termed the computational basis, by
$|n\rangle,\;n=0,1,..d-1;\;|n+d\rangle=|n\rangle$, these operators are defined by:
\begin{equation}\label{def}
Z|n\rangle=\omega^{n}|n\rangle;\;X|n\rangle=|n+1\rangle,
\end{equation}
with $\omega=e^{2\pi i/d}$. They form a complete set, i.e. only a multiple of the identity
commutes with both $X,\;Z.$ We shall briefly outline a method utilizing these operators
(due in the main to \cite{tal}) to construct the $d+1$ MUB for a d dimensional Hilbert
space with d being an odd prime. This review will be of help in building our sets of
entangled states that we shall associate with these MUB. The computational basis vectors
spans the Hilbert space. All operators in this space are expressible in terms of the $d^2$
Schwinger-operators \cite{schwinger}:
\begin{equation}\label{SO}
X^mZ^l;\;m,l=0,1,...d-1.
\end{equation}
The operators ${X^mZ^l}$ whose number is obviously $d^2$ form an orthogonal basis for all
operators in the d dimensional Hilbert space,
\begin{equation}
{\rm
Tr}\left[X^mZ^l\left(X^{m'}Z^{l'}\right)^{\dagger}\right]\;=\;\delta_{m,m'}\delta_{l,l'}.
\end{equation}
This follows from Eq.(\ref{def}) which implies the commutation formula
\begin{equation}\label{com1}
XZ=\omega ZX\;.
\end{equation}
Now, let us confine ourselves to cases wherein $m,l\;\in\ \mathbb{F}_d$ where $\mathbb{F}_d
$ is a d dimensional Galois field. In this case, we can relate Schwinger operators to MUB.
For this aim we group Schwinger operators  (\ref{SO}) into $d+1$ sets of $d-1$ orthogonal
commuting operators (which together with the identity operator form a complete operator
basis to Hilbert space). Each set of (commuting, orthogonal) operators defines a unique
vector basis in Hilbert space. All the $d+1$ sets of bases form an MUB set.

Let us first consider the case $m=0$ in Eq.~(\ref{SO}). Readily, the operators  $Z^l$ with
$l=0,..d-1$ ($l=0$ is, trivially, the identity operator) form one set of commuting and
orthogonal operators. This set is diagonalized in the computational basis (c.f.
Eq.~(\ref{def})). Next consider the case $m\neq 0$. In this case a unique inverse $m^{-1}$
is define on  $\mathbb{F}_d $ and thus we can rewrite the operators (\ref{SO}) as
\begin{eqnarray}\label{power}
X^mZ^l\;=\;\omega^{\nu}(XZ^b)^{m},\\\nonumber
\end{eqnarray}
where $b=l/m,\;\nu=-\frac b2 m(m-1)$ and $b=0,1..,d-1;\;\;m=1,...,d-1$. Thus, we have
associated the $d(d-1)$ of the above operators in the following manner
\begin{equation}
X^mZ^{bm}\;\sim\;(XZ^b)^m.
\end{equation}
That is,  these operators differ at most by a unimodular c number. Now, for a fixed b, we
have $(d-1)$ orthogonal and commuting operators. There are d distinct such sets, each
labelled by b $b=0,1,...d-1$ which are orthogonal
\begin{equation}
{\rm
Tr}\left[(XZ^b)^{m}\left((XZ^{b'})^{m'}\right)^\dagger\right]\;=d\;\delta_{b,b'}\delta_{m,m'},\;\;m,m'\ne\;0.
\end{equation}
When each b-labelled set of the d-1 orthogonal, commuting operators is supplemented with
the identity operator it constitutes  a set of d unitary, orthogonal and commuting
operators. And as such, defines a vector basis for Hilbert space: The d (orthonormal)
vectors that diagonalize these operators in the set. (We remark, in passing, that the basis
is defined up to a choice of a phase factor which does not affect the following results.)
For each (b-labelled) set there exist a a unique vectors basis. Here, we are designating
the vectors that form the basis by $|b;c\rangle$, where the index b labels the basis and c
that particular vector in the basis b $(b,c=0,1...d-1)$. The expressions for these states
in terms of the computational basis is \cite{tal},
\begin{equation}\label{mubstate}
|b;c\rangle\;=\;\frac1{\sqrt{d}}\sum_{n=0}^{d-1}\omega^{\frac b2 n(n-1)-cn}|n\rangle,
\end{equation}
The eigenvalues are $\omega^c$.

The importance of the classification of the operators in this manner arises once
considering the relation between the different sets of the vector basis.  One can readily
check that these bases  form MUB, i.e.,
\begin{eqnarray}
\langle b;c|b;c'\rangle\;&=&\;\delta_{c,c'},\nonumber \\
|\langle b',c'|b,c\rangle|\;&=&\;\frac1{\sqrt{d}},\;\;b\ne b',
\end{eqnarray}
Thus, these d distinct sets of bases plus the computational basis (which mutually unbiased
to all of these sets)  form the maximal number of MUB, that is d+1. The proof of the last
formula involves the well known \cite{schroeder} Gaussian sums.

Since Schwinger operators, Eq.~(\ref{SO}) form an operator basis for d-dimensional Hilbert
space, so do the set of operators $(XZ^b)^{m}$ together with $Z^b$
($b=0,..,d-1;m=1,..d-1$).
 Hence we may write an arbitrary density operator as
\begin{eqnarray}\label{rawrho}
\rho&=&\frac{1}{d}\Bigg(\sum_{m=1}^{d-1}\sum_{b=0}^{d-1}{\rm Tr}\left[\rho
(XZ^b)^m \right]\left((XZ^b)^m\right)^{\dagger} \nonumber \\
&+&\sum_{l=0}^{d-1}{\rm Tr}\left[\rho Z^l \right]\left(Z^l\right)^{\dagger}\Bigg).
\end{eqnarray}
It is convenient to rewrite this equation upon adding and subtracting $m=0$ terms:
\begin{equation}\label{rawrhoplus}
\rho=\frac{1}{d}\Bigg(\sum_{m=0}^{d-1}\sum_{b=0}^{d-1}{\rm Tr}\left[\rho (XZ^b)^m
\right]\left((XZ^b)^m\right)^{\dagger} -\Bbb{I} +\sum_{l=0}^{d-1}{\rm Tr}\left[\rho Z^l
\right]\left(Z^l\right)^{\dagger}\Bigg).
\end{equation}
We now utilize the spectral representation for each of the operators,
\begin{equation}
XZ^b=\sum_{m}|m,b\rangle\omega^m\langle b,m|\;\rightarrow Tr\rho(XZ^b)^m=\sum_{n}\langle
b,n|\rho|b,n\rangle \omega^{nm}.
\end{equation}
Hence,
\begin{equation}
\rho=\sum_{b,n}|b,n \rangle \langle b,n|\rho|b,n\rangle \langle b,n|+\sum_{n}|n\rangle
\langle n|\rho|n \rangle \langle n|-\Bbb{I}.
\end{equation}
Or,
\begin{equation}
\langle n'|\rho|n" \rangle=\sum_{b,n}\langle n'|b,m\rangle\langle
b,n|\rho|b,n\rangle\langle b,n|n"\rangle+\langle n'|\rho|n"\rangle \delta_{n',n"}
-\delta_{n',n"}.
\end{equation}

This is an expression for $\rho$  in terms of probabilities. The numbers $\langle
b;n|\rho|b;n \rangle$ correspond to probabilities to find the state $|b;n \rangle$ in
$\rho$ i.e. observable quantities. We see that, as shown in \cite{ivanovich}, the state of
the system, i.e. $\rho$, is reconstructed via the $d+1$ measurements. Each of these
measurements yields the $d-1$ independent probabilities outcomes (since the sum of the
probabilities add to one). This gives $(d+1)(d-1)=d^2-1$ numbers that determine the density
matrix. It should be noted that
this holds because the operators are non degenerate.\\

\section {Conclusions and Remarks}

An approach to classical tomography is given in close analogy with the quantum state
reconstruction scheme based on the Wigner representative function obtained via the Radon
transform. We then reviewed an alternative route for state reconstruction which is based on
the notion of mutual unbiased bases. The latter was shown to be applicable to the finite
dimensional Hilbert spaces eschewing thereby the somewhat sticky issue of finite
dimensional Wigner function. An overview of quantum ideas: measurable quantities, Wigner function
and mutual unbiased bases (MUB) is outline. \\
The analysis underscores the intriguing fact that to reconstruct a quantum state we require
the probabilities of all the phase space plane - not merely the probabilities along the
position and momentum axes as might be implied by (the wrong) positive reply to Pauli's
query posed at the introduction.

Acknowledgments: Informative discussions with  Professors J. Zak, A. Mann and O. Kenneth
are gratefully acknowledged.

\end{document}